\documentclass{parcfd}

\usepackage{nopageno}
\usepackage{graphicx}
\usepackage{amsmath}
\usepackage{amsfonts}
\usepackage{amssymb}
\usepackage{hyperref}
\usepackage{layout}
\usepackage{lmodern}
\usepackage{subcaption}
\usepackage[dvipsnames]{xcolor}

\title{HARD CONSTRAINT PROJECTION IN A PHYSICS INFORMED NEURAL NETWORK
}

\author{Miranda J. S. Horne$^{*}$, Peter K. Jimack$^{*}$, Amirul Khan$^{\dag}$, AND He Wang$^{\ddag}$}

\heading{Miranda J. S. Horne, Peter K. Jimack, Amirul Khan, and He Wang}

\address{$^{*\dag}$University of Leeds\\
$^{*}$School of Computing, $^{\dag}$School of Civil Engineering\\
Woodhouse Lane, Leeds, LS2 9JT, UK\\
scmho, p.k.jimack, a.khan@leeds.ac.uk, www.leeds.ac.uk 
\and
$^{\ddag}$University College London\\
Department of Computer Science\\
Gower Street, London, WC1E 6BT, UK\\
he\_wang@ucl.ac.uk, https://profiles.ucl.ac.uk/93306-he-wang}


\abstract{{In this work, we embed hard constraints in a physics informed neural network (PINN) which predicts solutions to the 2D incompressible Navier Stokes equations. We extend the hard constraint method introduced by Chen et al. (arXiv:2012.06148) 
from a linear PDE 
to a strongly non-linear PDE. 
The PINN is used to estimate the stream function and pressure of the fluid, and by differentiating the stream function we can recover an incompressible velocity field.
An unlearnable hard constraint projection (HCP) layer projects the predicted velocity and pressure to a hyperplane that admits only exact solutions to a discretised form of the governing equations.} 
}

\begin{document}

\section{INTRODUCTION}
Machine learning provides a promising framework to simulate fluid dynamics at a fraction of the computational cost of traditional numerical methods\cite{annualrevbrunton}. Furthermore, the incorporation of domain knowledge into a neural network can improve the prediction accuracy, increase the model’s explainability, and result in a neural network that is less reliant on training data.
Typically, the incorporation of the physical constraints into a neural network is only weakly enforced, for example, a PINN\cite{raissipinn} weakly enforces the governing equation by incorporating a penalty term (often the equation's residuals) into the loss function. 
In the cases where a physical constraint is strongly imposed, the enforced governing equation is often either linear\cite{chenhcp}, or an additional conservation law (such as the incompressibility constraint\cite{PhyCAE}).
In this paper we propose a method to strictly enforce the discretised form of a nonlinear partial differential equation, through projection, 
{inspired by the linear analogue used by Chen et al. in 2021\cite{chenhcp}.}

\section{PHYSICS INFORMED NEURAL NETWORK ({PINN})}
{We will consider} the 2D incompressible Navier Stokes equations (\ref{2dnse1},\ref{2dnse2},\ref{2dnseincom})\cite{firstcoursefluiddynamics}. The solution of this system is given by the 2D instantaneous 
velocity 
($u(x,y,t)$, $v(x,y,t)$) and the pressure ($p(x,y,t)$) of the fluid, where $x,y$ are the spatial coordinates, and $t$ is the temporal coordinate. 
The kinematic viscosity of the fluid is given by $\nu$, and the constant density by $\rho$.
\begin{equation}
0 = \frac{\partial u }{\partial t} + \left(
u \frac{\partial u}{\partial x} +
v \frac{\partial u}{\partial y} \right)
- \nu \frac{\partial^2 u}{\partial x^2}
- \nu \frac{\partial^2 u}{\partial y^2}
+ \frac{1}{\rho} \frac{\partial P}{\partial x}
\label{2dnse1} \end{equation}
\begin{equation}
0 = \frac{\partial v }{\partial t} + \left(
u \frac{\partial v}{\partial x}+
v \frac{\partial v}{\partial y} \right)
- \nu\frac{\partial^2 v}{\partial x^2}
- \nu\frac{\partial^2 v}{\partial y^2}
+\frac{1}{\rho}\frac{\partial P}{\partial y}
\label{2dnse2}\end{equation}
\begin{equation}
0= \frac{\partial u}{\partial x} + \frac{\partial v}{\partial y}
\label{2dnseincom}
\end{equation}

{The network proposed is a feed forward neural network (FFNN) 
that takes as input the coordinates of the system $(x,y,t)$ 
and outputs a prediction of the stream function ($\psi$) and the pressure ($p$) at those coordinates. The velocity components are defined as $u = {\partial \psi}/{\partial y}$, $v = {-\partial\psi}/{\partial x}$, using the automatic differentiation\cite{ADcite} (AD) built into the network. This method strictly imposes the incompressiblity (\ref{2dnseincom}) of the system.}

{As just coordinates alone would be insufficient to predict a unique solution, we use $N$ ground truth solutions ($u_i,v_i$) to anchor the model's predictions ($\hat{u}_i$, $\hat{v}_i$) to our test case.
The data error (DE, \ref{de_v1}) measures the distance between the ground truth solutions and the network predictions.}
We calculate the data error only on the velocity as in practice it would be significantly more difficult to measure the pressure of a fluid through the domain than the velocity when creating the ground truth data set.

\begin{equation}
\mathrm{DE}= 
\frac{1}{N}\sum_{i=1}^{N}(u_i- \hat{u}_i)^2 +
\frac{1}{N}\sum_{i=1}^{N}(v_i- \hat{v}_i)^2
\label{de_v1}    \end{equation}

Training the FFNN using only the DE would neglect the other knowledge we have of the system, the governing PDE.
A PINN\cite{raissipinn} appends the loss function to include some measure of the prediction's deviation from the governing equations, the physics error (PE, \ref{pe_v1}).
The Navier Stokes equation residuals (NSER) are found by evaluating the RHS of (\ref{2dnse1}) and (\ref{2dnse2}) using 
AD for $M$ of the network predictions at locations ($x_i,y_i,t_i$), denoted $r^x_i$ and $r^y_i$, for $i \in \{1,...,M\}$.
\begin{equation}
\mathrm{PE} =\frac{1}{M}\sum_{i=1}^{M}
(r^x_i)^2+\frac{1}{M}\sum_{i=1}^{M}(r^y_i)^2
\label{pe_v1}    \end{equation}

In some PINN literature (e.g.\cite{raissipinn}) the error from the initial condition and boundary conditions are incorporated {in the loss function} as separately weighted penalty terms.
In the models presented here, 
the values on the spatial and temporal boundaries are
incorporated into the data error, such that the loss function is defined only as a weighted sum of the DE and PE.

\section{HARD CONSTRAINT PROJECTION ({HCP})}
\begin{figure}[t]
\centering
\includegraphics[width=13cm]{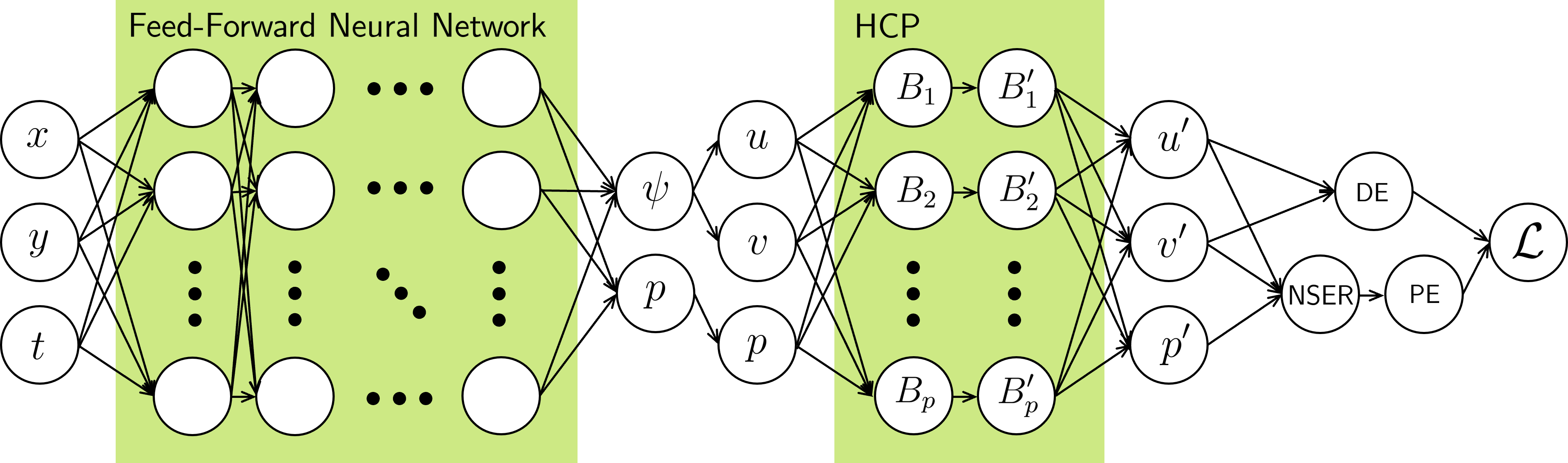}
\caption{{HCP-PINN} architecture. The left box is the {FFNN} with learnable weights and biases, and the right box is the unlearnable {HCP}.}
\label{HCPPINN2}
\end{figure}

{The HCP-PINN has the same learnable network and loss function as a PINN, 
but between the initial prediction of the solution and the loss evaluation there is an unlearnable hard constraint projection layer (see figure \ref{HCPPINN2})}.
Following the methodology used by Chen et al.\cite{chenhcp}, the governing equations (\ref{2dnse1},\ref{2dnse2})
 are discretised with a central finite difference scheme for the spatial derivatives, and a backwards finite difference scheme for the temporal derivatives. 
{The discretised forms of the governing equations are then decomposed into two matrices, the constraint matrix $A$ (containing all the constant terms) and the prediction matrix $B$ (containing all of the transient terms) such that multiplying $AB$ recovers the discretised governing equations (\ref{AB2}).}

{Using tools from linear algebra we can define the projection matrix as $ P = I - A^T(AA^T)^{-1}A$. 
When an arbitrary prediction matrix ($B$) is multiplied by the projection matrix, $PB=B'$, the outcome ($B'$) is the closest point to $B$ that lies on the hyperplane $AB=0$. Thus, the predictions after projection satisfy the discretised form of the governing equations exactly.}
For a proof of this, please see the appendix of the paper by Chen et al.\cite{chenhcp}.
{In practice, each prediction matrix contains only the values from one finite difference stencil, so each of the $B_i$ in figure \ref{HCPPINN2} corresponds to each of the input coordinate tuples $(x,y,t)$.} 
\begin{equation} \label{AB2} AB = \begin{bmatrix}
{1/ \Delta t} \\ {1/ \Delta t} \\
{1/ \Delta x} \\ {1/ \Delta x} \\ 
{1/ \Delta x} \\ {1/ \Delta y} \\
{1/ \Delta y} \\ {1/ \Delta y} \\ 
{1/ (\Delta x)^2} \\ {1/ (\Delta y)^2}
\end{bmatrix}^T 
\begin{bmatrix}
{u} &{v} \\
-u_{-\Delta t} &-v_{-\Delta t} \\
\frac{1}{2}u(u_{+\Delta x}-u_{-\Delta x})&\frac{1}{2}u(v_{+\Delta x}-v_{ -\Delta x})\\
{\tfrac{1}{2\rho}P_{+\Delta x}}&0\\
{\tfrac{-1}{2\rho}P_{-\Delta x}}&0\\
\frac{1}{2}v(u_{+\Delta y}-u_{-\Delta y})&
\frac{1}{2}v(v_{+\Delta y}-v_{ -\Delta y})\\
0&{\tfrac{1}{2\rho}P_{+\Delta y}}\\
0&{\tfrac{-1}{2\rho}P_{-\Delta y}}\\
\nu(-u_{+\Delta x} + 2u - u_{-\Delta x}) &\nu(-v_{+\Delta x} + 2v - v_{-\Delta x}) \\
\nu(-u_{+\Delta y} + 2u - u_{-\Delta y}) &\nu(-v_{+\Delta y} + 2v - v_{-\Delta y}) 
\end{bmatrix}
\end{equation}

\section{PROVISIONAL RESULTS}
\begin{figure}[t]
\centering
\includegraphics[width=\linewidth]{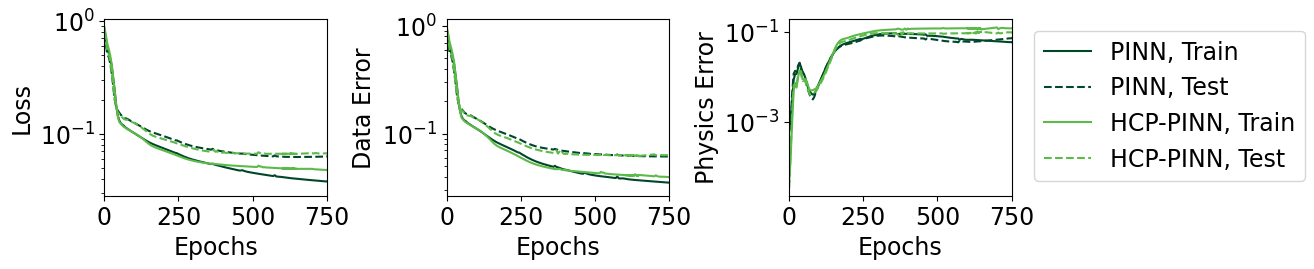}
\caption{Graphs comparing the value of the loss, data error, and physics error, on the test and training set, during the training of both models.
}
\label{results}
\end{figure}

To test the HCP-PINN, two models were trained, a PINN with no HCP layer, and a HCP-PINN as depicted in figure \ref{HCPPINN2}.
Both models had 6 hidden layers with 50 neurons each, and were optimised with default settings of the Adam\cite{kingmaadam} optimiser with a loss function defined as $\mathcal{L}=0.9\mathrm{DE}+0.1\mathrm{PE}$. The hyperbolic tangent was used as the activation function. 
{The discretisation used in the hard constraint projection has values $\Delta x=\Delta y=\Delta t=0.01$.}

The models were trained and tested on one dataset of periodic vortex shedding past a bluff body. 
The data domain is defined by $t \in [0,10]$, $x \in [1,8]$, and $y \in [-2,2]$, downstream of the bluff body and with the wake fully realised. The training dataset was random uniformly sampled across the domain with $N=500$ ground truth data points ($230$ of which lie on the spatial and temporal boundaries) and $M=1000$ physics collocation points. 
The test dataset is selected on a grid with $t = \{1.6,4.8,8.0\}$ and $\Delta x = 0.7070, \Delta y = 0.8164$ which both the test DE and the test PE are evaluated on ($N_{test}=M_{test}=150$).

\begin{figure}[t]
\centering
\includegraphics[width=0.9\linewidth]{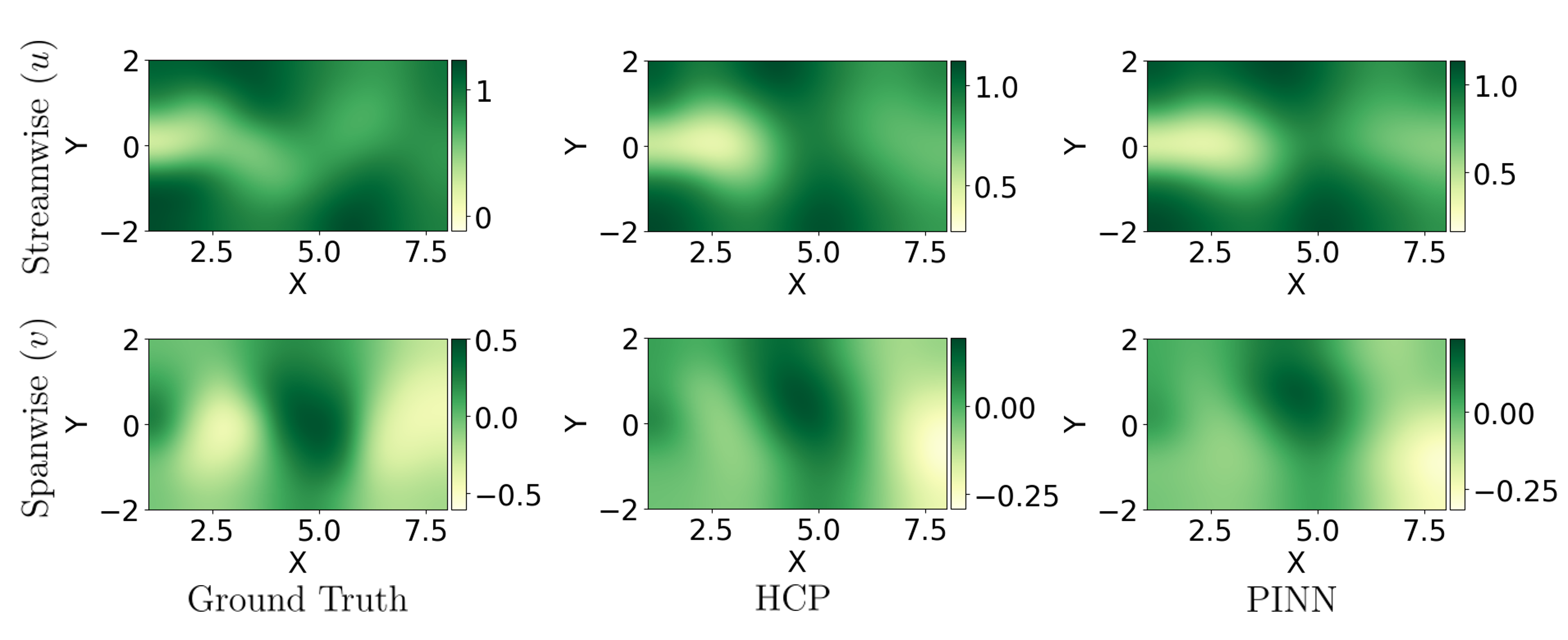}
\caption{Ground truth velocity field, and corresponding model predictions at epoch 750, of the velocity field at $t=4.8$.}
\label{results_flow}
\end{figure}

The training trajectory of the performance of the models can be seen in figure \ref{results}, and the predicted velocity fields at $t=4.8$ are displayed alongside the ground truth in figure \ref{results_flow}. 
We see that the training trajectories of the two models follow a similar shape, implying that the HCP-PINN optimises in a similar manner to the vanilla PINN. 
This is supported by the predictions in figure \ref{results_flow}, where the two models predict similar flow fields that qualitatively represent the key features of the flow.
We would not expect the physics error of the HCP-PINN to be exactly zero, as the governing equation is only exactly obeyed in its discretised form, locally. Unfortunately, we also find that the physics error associated with the HCP-PINN is not consistently lower than for the PINN, which was one of the motivations behind this implementation. 
 
An established issue in the literature on the training of PINNs is the imbalance between loss function terms\cite{wang2023expertsguidePINN}, and it was hoped by the authors that we would find the HCP-PINN less sensitive to the hyperparameter $w$ in the function $\mathcal{L}=(w)\mathrm{DE}+(1-w)\mathrm{PE}$. 
We had also anticipated that the HCP-PINN would potentially be less dependent on the quantity and sampling strategy for the physics collocation points, especially since the calculation of the residual at these points is a computational bottle-neck for both models.
We report that the HCP-PINN and PINN appear to respond equally sensitively from our studies into this (which have been omitted from the extended abstract for brevity). 
The authors suspect that the hard constraint projection, despite requiring greater computational resources, has a minimal effect on the predictions made during training, as implied by the similar training trajectories and predictions in figures \ref{results} and \ref{results_flow}.

The authors intend to look into modifying the implementation of the HCP-PINN, with the intention of more favourable results. 
We will look into the execution of the HCP, which uses a low order and potentially unstable discretisation, and a non-unique decomposition of the governing equation. We will also look at employing established machine learning methods such as batch training and transfer learning with the goal of fully exploring the potential of this method.
We finally note that Chen et al.\cite{chenhcp} investigated only one linear PDE when originally proposing this method for hard constraint projection, and it is possible that the HCP method is only appropriate for a subset of PDEs, such as linear or weakly non-linear PDEs.

\section{CURRENT WORK}

We aim to refine the HCP-PINN further. The directions stated in the previous section will be our immediate goals, however this model also has the potential to embed the boundary conditions strictly through the use of ghost cells.
Additionally, we would like to investigate the model's robustness to noise and outliers, generalisability to other flow regimes, and extrapolation capabilities given only boundary and initial conditions.

\bibliographystyle{ieeetr}
\bibliography{MJSH_Bib}

\end{document}